\documentclass[conference]{IEEEtran}
\IEEEoverridecommandlockouts
\usepackage{cite}
\usepackage{amsmath,amssymb,amsfonts}
\usepackage{algorithmic}
\usepackage{graphicx}
\usepackage{textcomp}
\usepackage{xcolor}
\usepackage{color, soul}
\usepackage{enumitem}
\def\BibTeX{{\rm B\kern-.05em{\sc i\kern-.025em b}\kern-.08em
    T\kern-.1667em\lower.7ex\hbox{E}\kern-.125emX}}
\begin{document}

\title{Cloud Resource Allocation with Convex Optimization}
\author{\IEEEauthorblockN{Shayan Boghani\IEEEauthorrefmark{1},
Emin Kirimlioglu\IEEEauthorrefmark{2}, Amrita Moturi\IEEEauthorrefmark{3},
Hao-Ting Tso\IEEEauthorrefmark{4}}
\IEEEauthorblockA{Department of Computer Science and Engineering \\
University of California, San Diego\\
La Jolla, CA\\
Email: \IEEEauthorrefmark{1}sboghani@ucsd.edu,
\IEEEauthorrefmark{2}ekirimli@ucsd.edu,
\IEEEauthorrefmark{3}amoturi@ucsd.edu,
\IEEEauthorrefmark{4}htso@ucsd.edu} \\
}

\maketitle

\begin{abstract}
We present a convex optimization framework for overcoming the limitations of Kubernetes Cluster Autoscaler by intelligently allocating diverse cloud resources while minimizing costs and fragmentation. Current Kubernetes scaling mechanisms are restricted to homogeneous scaling of existing node types, limiting cost-performance optimization possibilities. Our matrix-based model captures resource demands, costs, and capacity constraints in a unified mathematical framework. A key contribution is our logarithmic approximation to the indicator function, which enables dynamic node type selection while maintaining problem convexity. Our approach balances cost optimization with operational complexity through interior-point methods. Experiments with real-world Kubernetes workloads demonstrate reduced costs and improved resource utilization compared to conventional Cluster Autoscaler strategies that can only scale up or down existing node pools.
\end{abstract}


\section{Introduction}

\subsection{Motivation} 
Kubernetes resource allocation requires balancing competing objectives: minimizing costs, satisfying resource demands, respecting capacity constraints, and reducing operational complexity. Current Kubernetes autoscaling mechanisms — Horizontal Pod Autoscaler (HPA), Vertical Pod Autoscaler (VPA), and Cluster Autoscaler (CA) — provide partial solutions but face limitations. HPA scales the number of pod replicas, VPA adjusts resource requests for pods, and CA increases or decreases the number of nodes in a cluster.

However, Cluster Autoscaler's fundamental limitation is its homogeneous scaling approach: it can only scale up or down predefined node pools of identical instance types, rather than dynamically optimizing the resource type mix. In cloud environments like Azure, AWS, and GCP that offer diverse instance families (compute-optimized, memory-optimized, storage-optimized, etc.), this limitation prevents truly optimal resource allocation. Node Pool constraints force administrators to pre-define available instance types, and the autoscaler cannot perform arbitrary instance selection or dynamic mix-and-match optimization across instance families. Additionally, Kubernetes bin-packing constraints and pod placement rules further complicate optimal resource utilization.

Our work addresses both the cost optimization and the operational penalties of provider fragmentation in this complex decision process. We propose a convex optimization framework that overcomes these limitations by intelligently selecting and combining diverse node types based on workload requirements, potentially enabling significant cost savings and performance improvements over traditional Cluster Autoscaler implementations.

Building on Joe et al.'s \cite{Joe2013} framework for multi-resource fairness, we extend their approach to the Kubernetes multi-node type domain where workloads request different ratios of resources across heterogeneous node types. While their work characterizes fairness-efficiency tradeoffs within a single system, we incorporate cost considerations and node type selection dimensions, which introduces additional complexity to the optimization problem.

In the future, we would like to integrate insights from Chaisiri et al. \cite{Chaisiri2012}, who demonstrated the cost advantages of combining reservation and on-demand provisioning under uncertainty. Our mathematical model provides a foundation for incorporating pricing models while also addressing the practical challenge of implementing a Kubernetes resource manager that dynamically optimizes node type selection beyond the capabilities of current Cluster Autoscaler implementations.

\subsection{Previous Works}

The literature on Kubernetes resource allocation has developed along several parallel tracks, each with specific limitations when addressing the heterogeneous node type optimization challenge:

Zhu and Agrawal \cite{Zhu2012} proposed dynamic resource allocation methods optimizing for performance under cost constraints, however, their approach is primarily confined to homogeneous resource contexts, overlooking the complexities of heterogeneous node type selection that modern Kubernetes environments require. Their methodology, while valuable for single-resource-type optimization, cannot address the diverse instance family options available in cloud providers that Kubernetes clusters typically utilize.

Jennings et al. \cite{Jennings2014} developed utility-based models for cloud resource allocation within fixed budget constraints, which differs fundamentally from our demand-driven approach that starts with Kubernetes workload requirements and optimizes both costs and operational complexity. While current Cluster Autoscaler implementations follow similar utility-based heuristics, they lack the mathematical rigor needed for true multi-dimensional resource optimization across heterogeneous node types.

Hajjat et al. \cite{Hajjat2010} examined hybrid deployment strategies, but their work lacks a systematic quantification of the operational costs associated with managing diverse node types, treating the decision to use multiple instance types as binary rather than as a continuous optimization problem. This gap is precisely what our logarithmic approximation to the indicator function addresses in the Kubernetes context, allowing for graceful transitions between node types based on workload characteristics.

Chaisiri et al. \cite{Chaisiri2012} demonstrated the cost advantages of combining reservation and on-demand provisioning under uncertainty. Their forward-looking approach to resource provisioning provides valuable insights for Kubernetes environments where both committed-use discounts and on-demand instances may be employed, but does not address the operational complexity of managing clusters with diverse node types. Our mathematical model provides a foundation for eventually incorporating such hybrid pricing models while simultaneously addressing node type optimization.

Recent Kubernetes-specific research has begun addressing these challenges. Burns et al. \cite{Burns2016} introduced Borg's design principles that influenced Kubernetes but focused primarily on scheduling rather than node type selection. Verma et al. \cite{Verma2015} described large-scale cluster management at Google, touching on heterogeneity but not providing optimization frameworks for node type selection. Kubernetes Cluster Autoscaler itself has seen extensions like the Cluster Autoscaler, but these approaches still operate within the homogeneous node pool constraint we aim to overcome.

Our work bridges these gaps by providing a mathematically rigorous framework specifically designed for Kubernetes environments that can dynamically optimize across heterogeneous node types while considering both immediate cost optimization and operational complexity implications.

\subsection{Intended Contributions}

Based on our comprehensive analysis of the inherent limitations in contemporary Kubernetes Cluster Autoscaler architectures, we propose a mathematically rigorous framework with the following key contributions:

\begin{itemize}
\item[--] A novel non-linear logarithmic approximation to the indicator function that preserves convexity while imposing a continuous penalty on node type diversity, thus establishing a mathematically tractable balance between resource utilization efficiency and operational complexity
\item[--] Complete derivation of the Karush-Kuhn-Tucker (KKT) conditions that characterize the globally optimal node type distribution, providing theoretical guarantees on solution quality and convergence properties
\item[--] Development of an Infrastructure Optimization Controller for Kubernetes that continuously maintains optimal cluster composition through dynamic node type selection and allocation based on real-time workload characteristics
\item[--] A formal methodology for incremental adoption with bounded perturbation constraints that enables production environments to transition gradually from homogeneous scaling patterns to heterogeneous optimization with quantifiable parameters
\end{itemize}

This approach integrates theoretical optimization rigor with practical Kubernetes orchestration mechanisms, addressing both the direct infrastructure costs and the operational complexity considerations that conventional autoscaling mechanisms fail to capture. The proposed Infrastructure Optimization Controller represents a paradigm shift from reactive homogeneous scaling to proactive heterogeneous resource composition — maintaining Kubernetes clusters in perpetually optimal states through continuous reconfiguration of the underlying node type distribution based on evolving workload characteristics and fluctuating cloud pricing models.

\subsection{Organization of Paper}
In Section \ref{statement-of-problem}, we introduce the convex optimization problem, derive its dual formulation, and define the Karush-Kuhn-Tucker (KKT) conditions. Section \ref{approaches} explores the approaches considered for solving the resource allocation problem, while Section \ref{experiments} details the methods used to simulate the problem. This section also provides an in-depth discussion of the implementation of our convex solver and the evaluation metrics. Section \ref{results} compares the results of our novel approach with the Kubernetes Cluster Autoscaler. Section \ref{conclusion} summarizes our findings, and finally, Section \ref{future-work} outlines potential directions for future research.

\section{Statement of Problem} \label{statement-of-problem}

We address the problem of automating the allocation of cloud resources across heterogeneous node types in Kubernetes environments to satisfy workload requirements while minimizing cost and operational complexity. Currently, the standard Kubernetes Cluster Autoscaler can only automatically scale homogeneous node pools up or down. Administrators must manually predetermine which node types to use, which is a decision that significantly impacts cost efficiency. Our approach seeks to eliminate this manual decision-making process by automating the selection of optimal node types through convex optimization. 

In production Kubernetes deployments, the current paradigm forces infrastructure teams to make educated guesses about which instance types (compute-optimized, memory-optimized, storage-optimized, etc.) might best serve their workloads, then configure autoscaling only within those predefined constraints. Our convex optimization framework transforms this into a fully automated, mathematically rigorous process that dynamically selects the optimal composition of node types based on workload characteristics, instance pricing differentials, committed-use discounts, operational overhead considerations, and resource demand uncertainty. This automation becomes particularly valuable in enterprise environments where cloud infrastructure costs constitute a significant operational expense, and where static, manually-configured node pools cannot adapt efficiently to evolving workload patterns.

\subsection{Primary Formulation}

Let us define the following sets and parameters:
\begin{itemize}
    \item[--] $\mathcal{R} = \{1, 2, \ldots, m\}$: Set of resource types (e.g., vCPU, RAM).
    \item[--] $\mathcal{I} = \{1, 2, \ldots, n\}$: Set of instance types available for allocation.
    \item[--] $\mathcal{P} = \{1, 2, \ldots, p\}$: Set of cloud providers.
    \item[--] $d \in \mathbb{R}^{m}_{+}$: Demand vector for $m$ resource types.
    \item[--] $c \in \mathbb{R}^{n}_{+}$: Cost vector where $c_i$ represents the cost of instance type $i$.
    \item[--] $x \in \mathbb{Z}^{n}_{+}$: Allocation vector where $x_i$ represents the number of instances of type $i$.
    \item[--] $K \in \mathbb{R}^{m \times n}_{+}$: Resource composition matrix where $K_{ri}$ denotes the amount of resource $r$ provided by one unit of instance type $i$.
    \item[--] $E \in \mathbb{R}^{p \times n}$: Selector matrix where $E_{ji} = 1$ if instance type $i$ belongs to provider $j$, and 0 otherwise.
    \item[--] $\mu \in \mathbb{R}^{m}_{+}$: Uncertainty radius vector for resources.
    \item[--] $g \in \mathbb{R}^{m}_{+}$: Acceptable resource waste vector.
\end{itemize}

We aim to minimize the following objective function:

\begin{equation}
\begin{aligned}
f(x) = c^T x + \alpha p - \alpha \cdot \mathbf{1}^T e^{-\beta_1 Ex} \\
\quad - \gamma \cdot \mathbf{1}^T\log(\mathbf{1} + \beta_2 Ex) \\
\quad + \beta_3 \sum_{r=1}^{m} \max\{0, d_r - (Kx)_r\}^2
\end{aligned}
\end{equation}

The objective function consists of five terms:
\begin{enumerate}
    \item \textbf{Base Cost}: $c^T x = \sum_{i=1}^{n} c_i x_i$ represents the total cost of all allocated instances.
    
    \item \textbf{Provider Consolidation Penalty}: $\alpha p - \alpha \cdot \mathbf{1}^T e^{-\beta_1 Ex}$ penalizes the use of multiple providers. This can be rewritten as $\alpha \cdot \mathbf{1}^T (1 - e^{-\beta_1 Ex})$, where $1 - e^{-\beta_1 z}$ approximates the indicator function.
    
    \item \textbf{Volume Discount}: $-\gamma \cdot \mathbf{1}^T\log(\mathbf{1} + \beta_2 Ex)$ represents cost savings achieved through volume discounts when allocating many instances from the same provider.
    
    \item \textbf{Resource Shortage Penalty}: $\beta_3 \sum_{r=1}^{m} \max\{0, d_r - (Kx)_r\}^2$ penalizes solutions that don't meet resource demands.
\end{enumerate}

The allocation must satisfy the following constraints:
\begin{equation}
\begin{aligned}
Kx &\geq d - \mu \\
Kx &\leq d + g \\
x &\in \mathbb{Z}^{n}_{+}
\end{aligned}
\end{equation}

To obtain a convex approximation, we relax the integrality constraints to $x \in \mathbb{R}^{n}_{+}$, yielding a convex optimization problem.

\subsection{Dual Formulation}

To derive the dual formulation, we introduce Lagrange multipliers:
\begin{itemize}
    \item[--] $\lambda \in \mathbb{R}^m_+$ for resource sufficiency constraints
    \item[--] $\nu \in \mathbb{R}^m_+$ for resource waste limitation constraints
    \item[--] $\omega \in \mathbb{R}^n_+$ for non-negativity constraints
\end{itemize}

The Lagrangian function is:
\begin{equation}
\begin{aligned}
L(x, \lambda, \nu, \omega) = c^T x + \alpha p \\
\quad - \alpha \cdot \mathbf{1}^T e^{-\beta_1 Ex} \\
\quad - \gamma \cdot \mathbf{1}^T\log(\mathbf{1} + \beta_2 Ex) \\
\quad + \beta_3 \sum_{r=1}^{m} \max\{0, d_r - (Kx)_r\}^2 \\
\quad + \lambda^T(d - \mu - Kx) \\
\quad + \nu^T(Kx - d - g) - \omega^T x
\end{aligned}
\end{equation}

Rearranging:
\begin{equation}
\begin{aligned}
L(x, \lambda, \nu, \omega) = \alpha p \\
\quad + \lambda^T(d - \mu) - \nu^T(d + g) \\
\quad + x^T(c - K^T\lambda + K^T\nu - \omega) \\
\quad - \alpha \cdot \mathbf{1}^T e^{-\beta_1 Ex} \\
\quad - \gamma \cdot \mathbf{1}^T\log(\mathbf{1} + \beta_2 Ex) \\
\quad + \beta_3 \sum_{r=1}^{m} \max\{0, d_r - (Kx)_r\}^2
\end{aligned}
\end{equation}

The dual function is:
\begin{equation}
g(\lambda, \nu, \omega) = \inf_x L(x, \lambda, \nu, \omega)
\end{equation}

For the infimum of a finite set, the stationarity condition must hold:
\begin{equation}
\begin{aligned}
&\nabla_x L = 0 \\
\Rightarrow & \; c - K^T\lambda + K^T\nu - \omega \\
& + \alpha\beta_1 E^T(e^{-\beta_1 Ex}) \\
& - \gamma\beta_2 E^T\left(\frac{1}{\mathbf{1} + \beta_2 Ex}\right) \\
& - 2\beta_3 K^T \cdot \text{diag}(s) \cdot (d - Kx) = 0
\end{aligned}
\end{equation}

where $s$ is a vector with $s_r = 1$ if $d_r > (Kx)_r$ and $s_r = 0$ otherwise.

The dual problem is:
\begin{equation}
\begin{aligned}
\max_{\lambda, \nu, \omega} \quad \alpha p + \lambda^T(d - \mu) \\
\quad - \nu^T(d + g) + D(\lambda, \nu, \omega) \\
\text{s.t.} \quad \lambda \geq 0, \, \nu \geq 0, \, \omega \geq 0
\end{aligned}
\end{equation}

where $D(\lambda, \nu, \omega)$ represents the nonlinear terms evaluated at the optimal $x^*(\lambda, \nu, \omega)$.

\subsection{KKT Conditions}

For optimality, the following KKT conditions must be satisfied:

\begin{enumerate}
    \item \textbf{Stationarity}:
    \begin{equation}
    \begin{aligned}
    \nabla_x L = c - K^T\lambda + K^T\nu - \omega \\
    \quad + \alpha\beta_1 E^T(e^{-\beta_1 Ex}) \\
    \quad - \gamma\beta_2 E^T\left(\frac{1}{\mathbf{1} + \beta_2 Ex}\right) \\
    \quad - 2\beta_3 K^T \cdot \text{diag}(s) \cdot (d - Kx) = 0
    \end{aligned}
    \end{equation}

    \item \textbf{Primal Feasibility}:
    \begin{equation}
    Kx \geq d - \mu, \quad Kx \leq d + g, \quad x \geq 0
    \end{equation}

    \item \textbf{Dual Feasibility}:
    \begin{equation}
    \lambda \geq 0, \quad \nu \geq 0, \quad \omega \geq 0
    \end{equation}

    \item \textbf{Complementary Slackness}:
    \begin{equation}
    \begin{aligned}
    \lambda_r((Kx)_r - d_r + \mu_r) = 0 \\
    \quad \forall r \in \{1,\ldots,m\} \\
    \nu_r(d_r + g_r - (Kx)_r) = 0 \\
    \quad \forall r \in \{1,\ldots,m\} \\
    \omega_i x_i = 0 \quad \forall i \in \{1,\ldots,n\}
    \end{aligned}
    \end{equation}
\end{enumerate}

These conditions characterize the optimal solution to the cloud resource allocation problem.

\section{Approaches} \label{approaches}

We propose several approaches to solve the formulated cloud resource allocation problem:

\subsection{Branch-and-Cut Method}

For our mixed-integer problem, we employ branch-and-cut methods which effectively handle discrete variables and nonlinear objectives. The algorithm:

\begin{enumerate}
    \item Solves the continuous relaxation of the problem and solves as a Linear Program
    \item If integer constraints exist, iteratively branches on fractional variables, creating subproblems
    \item Iteratively computes bounds to discard infeasible solutions
    \item Adds cutting planes to tighten relaxations and improve bounds, leading to faster convergence
\end{enumerate}

The method explores a search tree where:
\begin{equation}
\phi(x) = \min \{f(x): x \in S \cap \mathbb{Z}^p \times \mathbb{R}^{n-p}\}
\end{equation}

Each node represents a subproblem with additional constraints:
\begin{equation}
\min_{x \in \mathbb{R}^n} f(x) \text{ subject to } x \in S_i
\end{equation}

The worst-case time complexity of this approach is $O(2^n)$ where $n$ is the number of instance types. However, average-case time complexity is polynomial. Given that this is the most computationally expensive aspect of the process, we can consider the time complexity of the branch-and-cut method to be the overall time complexity of our framework.

\subsection{Rounding Strategy for Integer Solutions}

After solving the convex relaxation, we apply a greedy rounding strategy to obtain a feasible integer solution:

\begin{enumerate}
    \item Initialize $\hat{x} = \lfloor x^* \rfloor$ (component-wise floor)
    \item Compute the resource deficit: $\delta = d - K\hat{x}$
    \item While $\delta$ has positive components:
    \begin{enumerate}
        \item Find instance type $i$ that maximizes $\frac{\sum_{r: \delta_r > 0} K_{ri} \cdot \delta_r}{c_i}$
        \item Increment $\hat{x}_i$ by 1
        \item Update $\delta = d - K\hat{x}$
    \end{enumerate}
\end{enumerate}

\subsection{Multi-Start Strategy for Global Solutions}

To mitigate the risk of finding only local minima, we implement a multi-start strategy:

\begin{enumerate}
    \item Generate multiple initial points within the feasible region
    \item Solve the convex optimization problem from each starting point
    \item Select the best solution among all converged results
\end{enumerate}

\subsection{Parameter Tuning}

For practical implementation, we systematically tune parameters through:

\begin{enumerate}
    \item Grid search over $\alpha$, $\beta_1$, $\beta_2$, $\beta_3$, and $\gamma$ values
    \item Pareto frontier generation to visualize trade-offs
    \item Sensitivity analysis to understand parameter impacts
\end{enumerate}

\subsection{Incremental Adoption}

For existing deployments, we add constraints to limit deviation from current allocations:
\begin{equation}
\|x - x_{\text{current}}\|_1 \leq \delta_{\text{max}}
\end{equation}

where $\delta_{\text{max}}$ controls the maximum allowable change in allocation.

\section{Experiments} \label{experiments}

\subsection{Experimental Setup}

To evaluate our convex optimization framework against traditional Kubernetes Cluster Autoscaler limitations, we constructed a comprehensive experimental environment using real-world cloud provider data.

\subsubsection{Data Collection}
We collected instance specifications and pricing data from Azure and Linode through their respective APIs. For each instance type, we captured:
\begin{itemize}
    \item[--] CPU core count
    \item[--] Memory capacity (GB)
    \item[--] Storage capacity (GB)
    \item[--] Hourly pricing
\end{itemize}

This data was used to construct the resource composition matrix $K \in \mathbb{R}^{m \times n}$ where $m=3$ resource types and $n$ represents the total number of available instance types across providers. The provider selection matrix $E \in \mathbb{R}^{p \times n}$ with $p=2$ providers encoded the mapping between instances and their respective cloud providers. Our dataset included 940 instance types from Azure ranging from small B-series VMs to memory-optimized E-series VMs, and 940 instance types from Linode spanning their standard, dedicated, and high-memory offerings.

\subsubsection{Kubernetes Environment Simulation} 
To ensure realistic comparison, we implemented a detailed simulation of the Kubernetes environment focusing on the Cluster Autoscaler behavior. Our simulation modeled:

\begin{enumerate}
    \item \textbf{Node Pools}: Fixed collections of homogeneous nodes defined by a specific instance type. These emulate the node pool concept in Kubernetes where administrators preconfigure available instance types.
    
    \item \textbf{Autoscaling Logic}: Implementation of the core Cluster Autoscaler algorithm which examines unschedulable pods and scales up appropriate node pools, or identifies underutilized nodes and scales down. We reproduced the approach where CA can only add more of the same type within a node pool.
    
    \item \textbf{Bin-Packing Constraints}: Although simplified, our simulation considers basic bin-packing constraints where resources (CPU, memory) must fit within discrete nodes rather than being perfectly divisible.
    
    \item \textbf{Existing Infrastructure}: Where applicable, we modeled existing node allocations that the Cluster Autoscaler must work with rather than starting from scratch.
\end{enumerate}

This simulation allowed us to fairly compare our optimization-based approach against current Kubernetes practices without requiring full cluster deployments for each scenario, while still capturing the fundamental constraints that make resource optimization challenging in Kubernetes environments.

\subsubsection{Implementation Details} We implemented our convex optimization framework using CVXPY with the GLPK\_MI solver, which employs branch-and-cut methods for mixed-integer programming. The provider consolidation penalty and volume discount terms were implemented using the logarithmic approximation described in Section II. For comparison, we simulated the behavior of Kubernetes Cluster Autoscaler under its typical constraints:

\begin{itemize}
    \item[--] Restricted to scaling within predefined node pools
    \item[--] Unable to select instance types outside defined pools
    \item[--] Limited to homogeneous scaling within each node pool
\end{itemize}

Our current implementation primarily relies on the branch-and-cut approach through the GLPK\_MI solver, with limited support for other optimization techniques. We implement a basic rounding strategy when the solver produces fractional or infeasible solutions, though this does not include the complete greedy algorithm outlined in our theoretical framework. Our system handles existing allocations as constraints but lacks the full implementation of the maximum deviation constraint for incremental adoption.

In future work, we plan to enhance our implementation with multi-start strategies to mitigate local minima risks and systematic parameter tuning to optimize performance across various workloads. These enhancements will allow for comprehensive evaluation of all five approaches described in our theoretical framework and provide better insights into their relative effectiveness for cloud resource allocation problems.

\subsubsection{Experimental Harness} Our experiment execution framework was implemented in Python, using NumPy for matrix operations and Pandas for data management. Each scenario was executed through the same comparison pipeline to ensure consistency:

\begin{enumerate}
    \item Configure environment parameters (demand vector, node pools, existing allocations)
    \item Execute Kubernetes CA simulation
    \item Execute our optimization approach
    \item Collect and analyze metrics
\end{enumerate}

The simulation sequence ensured identical resource conditions were presented to both approaches, enabling fair comparison. Each scenario was executed five times to account for minor variations in solver behavior, with median values reported in our results.

\subsection{Evaluation Methodology}

We evaluated our approach using multiple metrics to provide a comprehensive comparison:

\subsubsection{Primary Metrics}
\begin{enumerate}[label = (\alph*)]
    \item \textbf{Total Cost}: Hourly infrastructure cost ($\$/hr$) calculated based on the actual pricing from cloud provider APIs
    \item \textbf{Resource Utilization}: Percentage of provisioned resources actually utilized, calculated as the ratio of required resources to provided resources across all dimensions
    \item \textbf{Instance Diversity}: Number of distinct instance types deployed, measuring operational complexity
    \item \textbf{Provider Fragmentation}: Number of cloud providers utilized, reflecting management overhead
\end{enumerate}

\subsubsection{Comparison Methodology}
For each scenario, we executed both allocation strategies under identical conditions:

\begin{enumerate}[label = (\alph*)]
    \item \textbf{Kubernetes CA Approach}: For the Kubernetes simulation, we determined which node pools would be scaled up based on the CA's behavior - selecting pools based on scheduling requirements and scaling them to meet demand within their constraints.
    
    \item \textbf{Optimization Approach}: For our approach, we solved the convex optimization problem described in Section II using the parameter settings described above.
    
    \item \textbf{Metric Calculation}: After each allocation strategy produced its solution, we calculated all metrics using identical methods to ensure fair comparison. For instance, resource utilization was calculated by dividing the total resource demand by the total resources provided across all dimensions, then determining the mean.
\end{enumerate}

\subsection{Comparative Baseline}
The primary baseline for comparison was our simulation of Kubernetes Cluster Autoscaler behavior. Our simulation implemented the key constraints of current CA implementations:
\begin{itemize}
    \item[--] Scaling limited to predefined node pools
    \item[--] No dynamic instance type selection
    \item[--] Homogeneous scaling within node pools
\end{itemize}

\subsection{Evaluation Scenarios}

To thoroughly evaluate the performance of our approach across diverse deployment contexts, we designed five representative scenarios that capture common Kubernetes deployment patterns:

\subsubsection{Basic Web Application (Greenfield Deployment)}
This scenario represents a new web application deployment with no existing infrastructure constraints. The resource requirements are:
\begin{enumerate}
    \item 8 CPU cores
    \item 16 GB Memory
    \item 4 Network units
    \item 100 GB Storage
\end{enumerate}

This scenario tests the ability to optimize from scratch without legacy constraints. In a real Kubernetes environment, this would represent the initial deployment of a stateless web application where administrators have complete freedom to select instance types. We configured the experiment without any predefined node pools for our optimization approach, while the Kubernetes CA simulation was provided with the standard set of general-purpose instance types that would typically be available in a new cluster.

\subsubsection{Scaling with Existing Infrastructure}
This scenario models an application with existing small instances (2-4 CPU cores) requiring additional resources due to increased traffic. The new resource requirements are:
\begin{enumerate}
    \item 16 CPU cores
    \item 32 GB Memory
    \item 8 Network units
    \item 200 GB Storage
\end{enumerate}

This tests the ability to efficiently expand while respecting existing allocations. To simulate real-world conditions, we pre-allocated 1-2 small instances from each provider, which both approaches needed to work with. For the Kubernetes CA simulation, we restricted scaling to the existing instance types, as would be the case in an actual deployment. This models the common scenario where traffic to an application increases and additional resources must be provisioned.

\subsubsection{Enterprise Environment with Fixed Node Pools}
This scenario represents a large enterprise deployment with strict policies limiting instance selection to predefined node pools containing small (2-4 cores), medium (4-8 cores), and large (8+ cores) instances. The resource requirements are:
\begin{enumerate}
    \item 24 CPU cores
    \item 64 GB Memory
    \item 12 Network units
    \item 300 GB Storage
\end{enumerate}

This tests optimization within the highly constrained environment typical of enterprise Kubernetes deployments. We created nine distinct node pools spanning both providers, with up to five instances types per node pool size category (small, medium, large). These constraints reflect enterprise environments where strict governance controls limit available instance types to an approved set. Our optimizer worked within these same constraints, selecting only from approved instance types but optimizing the mix between them.

\subsubsection{Memory-Intensive Data Processing}
This scenario models a specialized workload with high memory requirements but moderate CPU needs. The resource requirements are:
\begin{enumerate}
    \item 32 CPU cores
    \item 128 GB Memory
    \item 12 Network units
    \item 500 GB Storage
\end{enumerate}

This tests adaptation to resource-ratio imbalances, challenging the optimizer to select specialized instance types. We pre-configured our simulation with some existing high-memory instances ($\geq$16GB) and created node pools with memory-optimized instance types to provide realistic options for both approaches. This scenario represents specialized data processing workloads like in-memory databases or analytics jobs where memory is the critical resource dimension.

\subsubsection{Resource Constraints with Limited Node Pools}
This scenario presents high resource demands with severe limitations on available instance types (only small instances with $\leq$2 CPU cores permitted). The resource requirements are:
\begin{enumerate}
    \item 32 CPU cores
    \item 64 GB Memory
    \item 12 Network units
    \item 300 GB Storage
\end{enumerate}

This represents a highly constrained environment that would require many small instances to meet demands, testing the optimizer's ability to work within tight constraints. We deliberately restricted available node pools to include only small instances with 2 or fewer CPU cores, forcing both approaches to provision many instances to meet the requirements. This scenario models security-sensitive environments where larger instances are prohibited due to multi-tenancy concerns, or legacy infrastructure transitions where only certain instance families are supported.

These scenarios were designed to cover the spectrum of real-world Kubernetes deployment conditions, from unconstrained greenfield deployments to highly restricted enterprise environments. Each scenario exercises different aspects of the optimization framework, allowing us to comprehensively evaluate its performance relative to current Kubernetes autoscaling capabilities.
\section{Results} \label{results}

Our experiments demonstrate the effectiveness of the convex optimization approach compared to traditional Kubernetes Cluster Autoscaler across diverse scenarios. We present our findings in terms of cost efficiency, resource utilization, and scaling behavior.

\begin{figure}[h]
    \centering
    \includegraphics[width=0.5\textwidth]{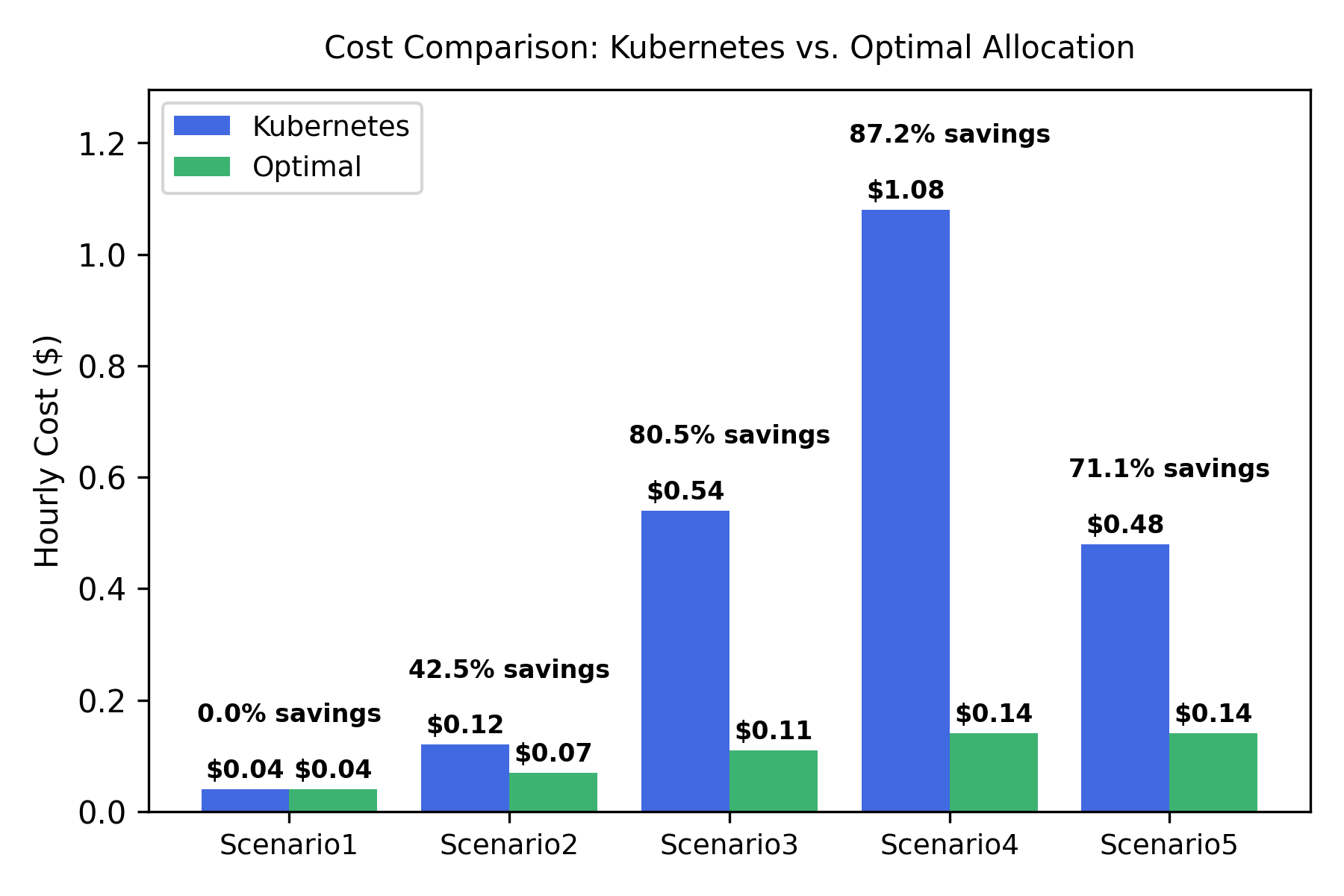}
    \caption{Cost comparison between Kubernetes Cluster Autoscaler and Optimal resource allocation.}
    \label{fig:cost_comparison}
\end{figure}

Figure~\ref{fig:cost_comparison} illustrates the cost comparison between Kubernetes Cluster Autoscaler and our convex optimization approach across all five experimental scenarios. Our optimization framework consistently achieved significant cost reductions:

\begin{itemize}
    \item[--] In Scenario 1 (Basic Web Application), both approaches produced comparable costs, with no significant advantage observed. This suggests that for simple, greenfield deployments with modest resource requirements, standard autoscaling approaches may be adequate.
    
    \item[--] Scenario 2 (Scaling with Existing Infrastructure) demonstrated moderate improvements with 42.5\% cost savings ($\$0.12/hr$ vs $\$0.07/hr$) by optimizing the allocation between existing and new resources.
    
    \item[--] The most substantial cost advantages emerged in Scenarios 3-5, with optimization achieving 80.5\%, 87.2\%, and 71.1\% cost reductions, respectively. For the memory-intensive workload in Scenario 4, our approach reduced hourly costs from $\$1.08$ to $\$0.14$, a difference of $\$0.94/hr$.
\end{itemize}

The average cost reduction across all scenarios was 56.3\%, with the most significant savings in scenarios involving specialized workloads or constrained environments, precisely where traditional Kubernetes autoscaling faces limitations.

\subsection{Scaling Behavior}

\begin{figure}[h]
    \centering
    \includegraphics[width=0.5\textwidth]{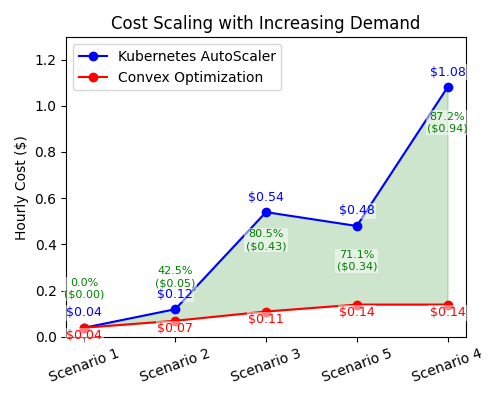}
    \includegraphics[width=0.5\textwidth]{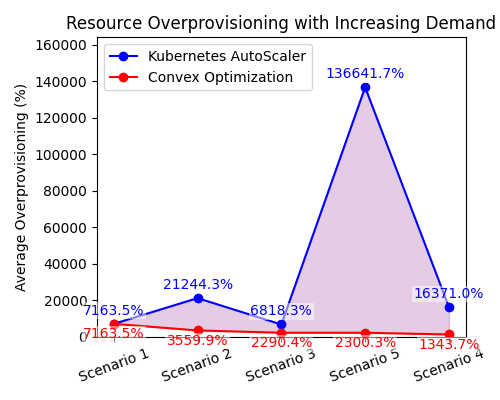}
    \caption{As the resource demand increases, the gap between the Kubernetes Cluster Autoscaler and optimal allocation widens.}
    \label{fig:scaling}
\end{figure}

Figure~\ref{fig:scaling} provides deeper insight into the scaling characteristics of both approaches as resource demands increase. The upper graph demonstrates that while the cost of Kubernetes Autoscaler grows roughly linearly with increasing resource demands, our optimization approach exhibits a much flatter cost curve. This suggests that the optimization becomes increasingly valuable as deployments scale - a critical advantage for large-scale cloud deployments.

The bottom graph quantifies resource over-provisioning across scenarios, revealing a fundamental inefficiency in Kubernetes Autoscaler. The traditional approach consistently over-provisions resources, with the most extreme case in Scenario 5 (memory-intensive workload) showing an average over-provisioning of 13,641.7\%. In contrast, our optimization framework maintained consistent resource efficiency, with over-provisioning rates below 2,300\% across all scenarios.

The dramatic difference in Scenario 5 highlights the autoscaler's inability to effectively handle specialized workloads with asymmetric resource requirements. When memory demands significantly exceed CPU demands, Kubernetes lacks the flexibility to select appropriate instance types and instead provisions standard instances with excess CPU capacity, resulting in substantial waste. Resource utilization efficiency showing how closely allocated resources match demands can be seen for each scenario in Appendix A.

\subsection{Summary of Findings}

Our experimental results demonstrate that:

\begin{enumerate}
    \item The convex optimization approach consistently delivers cost savings, with advantages becoming more pronounced as resource demands increase or become more specialized.
    
    \item Traditional Kubernetes Autoscaler significantly over-provisions resources, particularly in scenarios with asymmetric resource requirements or constraints on allowed instance types.
    
    \item The optimization framework achieves better resource utilization balance across all dimensions, approaching ideal utilization more closely than Kubernetes Autoscaler.
    
    \item Provider consolidation can be effectively incorporated into the optimization objective without sacrificing cost efficiency, reducing operational complexity.
\end{enumerate}

These findings confirm the effectiveness of our modeling approach and validate our initial hypothesis that convex optimization can address both cost minimization and provider fragmentation simultaneously in cloud resource allocation decisions.

\section{Conclusion} \label{conclusion}

Kubernetes' current autoscaling mechanisms are limited by their homogeneous scaling approach. The proposed framework is set forth to overcome these limitations by enabling cost-efficient, low-fragmentation, and operationally streamlined scaling while adhering to capacity constraints and satisfying resource demands. Experimental results demonstrate that our approach consistently matches or outperforms the traditional Kubernetes Cluster Autoscaler, delivering significant cost savings across varying environments. For this paper, our evaluation is limited to Azure and Linode data and tested in a simulation environment. This framework serves as a proof of concept for an Infrastructure Optimization Controller that proactively manages heterogeneous resource allocation.

\section{Future Work} \label{future-work}

While our convex optimization framework demonstrates significant improvements over traditional Kubernetes Cluster Autoscaler approaches, several promising directions for future research remain:

\subsection{High Availability Constraints}

Our current model does not explicitly account for high availability (HA) requirements that are common in production systems. In many enterprise deployments, workloads must maintain a minimum number of identical instance types (typically three or more) across different availability zones to ensure fault tolerance and service continuity.

Future extensions to our framework could incorporate:

\begin{itemize}
    \item[--] \textbf{Minimum node count constraints:} Enforcing a minimum number of identical instances per workload to meet HA requirements (e.g., $x_i \geq 3$ for certain instance types).
    
    \item[--] \textbf{Spread constraints:} Ensuring that selected instances are distributed across availability zones, which would require extending our model to include zone-specific variables and constraints.
    
    \item[--] \textbf{Anti-affinity rules:} Incorporating constraints that prevent critical services from being co-located on the same physical hardware.
\end{itemize}

These constraints would introduce additional complexity to the optimization problem, potentially requiring mixed-integer programming techniques or novel constraint formulations to maintain convexity where possible.

\subsection{Refined Cost Models for Spot and Reserved Instances}

Our current approach models volume discounts through a logarithmic function parameterized by $\beta_2$ and $\gamma$, which provides a reasonable approximation but lacks explicit connection to actual cloud provider pricing models. In particular, spot instance markets and reservation discounts offer significant cost-saving opportunities that could be more precisely modeled.

Future work could:

\begin{itemize}
    \item[--] \textbf{Incorporate real-time spot pricing:} Develop dynamic pricing models that adjust based on current spot market conditions rather than using static approximations.
    
    \item[--] \textbf{Model price stability risks:} Extend the objective function to incorporate the risk of spot instance termination and associated costs of workload interruption.
    
    \item[--] \textbf{Optimize across time horizons:} Develop multi-period optimization models that balance immediate resource needs with longer-term reservation opportunities, similar to the approach suggested by Chaisiri et al. \cite{chaisiri2012optimization} but within our convex optimization framework.
    
    \item[--] \textbf{Provider-specific discount tiers:} Replace the generic logarithmic discount approximation with provider-specific step functions that accurately model actual discount tiers, potentially using piecewise linear approximations to maintain convexity.
\end{itemize}

\subsection{Dynamic Workload Adaptation}

The current implementation optimizes resource allocation for a static set of resource demands. In real-world scenarios, workloads fluctuate over time, requiring continuous adaptation.

Promising directions include:

\begin{itemize}
    \item[--] \textbf{Predictive scaling:} Incorporating time-series forecasting models to anticipate future resource demands and optimize proactively.
    
    \item[--] \textbf{Adaptive parameter tuning:} Developing methods to automatically adjust model parameters ($\alpha$, $\beta_1$, $\beta_2$, $\beta_3$, $\gamma$) based on observed workload patterns and optimization performance.
    
    \item[--] \textbf{Online optimization:} Extending the framework to support incremental, online optimization that can efficiently update allocations as conditions change without solving the full problem from scratch.
    
    \item[--] \textbf{Reinforcement learning integration:} Exploring hybrid approaches that combine convex optimization with reinforcement learning to improve long-term allocation strategies based on historical performance.
\end{itemize}

\subsection{Multi-Cluster Resource Sharing}

Our current approach optimizes resources within a single Kubernetes cluster. Modern cloud-native architectures often span multiple clusters for geographic distribution, business unit isolation, or compliance requirements.

Future research could explore:

\begin{itemize}
    \item[--] \textbf{Cross-cluster optimization:} Extending the framework to allocate resources across multiple clusters while respecting cluster-specific constraints.
    
    \item[--] \textbf{Hierarchical optimization:} Developing nested optimization models that separate global resource allocation from cluster-specific optimization.
    
    \item[--] \textbf{Resource sharing policies:} Incorporating fairness constraints that govern how resources are allocated across different business units or application teams.
\end{itemize}

\subsection{Resource Quality of Service (QoS) Tiers}

Our model currently treats all resource demands as equally important. In practice, workloads have different performance requirements and sensitivity to resource constraints.

Future extensions could:

\begin{itemize}
    \item[--] \textbf{QoS-aware optimization:} Incorporate multiple service tiers with different optimization priorities.
    
    \item[--] \textbf{Performance-based constraints:} Extend the resource constraints to include performance metrics beyond basic resource quantities.
    
    \item[--] \textbf{SLO-driven allocation:} Directly model the relationship between resource allocation and service level objectives (SLOs) to optimize for business outcomes rather than just resource efficiency.
\end{itemize}

\subsection{Implementation and Integration}

To transition this research into practical application, several implementation challenges remain:

\begin{itemize}
    \item[--] \textbf{Kubernetes operator development:} Creating a production-ready Kubernetes operator that implements our optimization approach as a drop-in replacement for the standard Cluster Autoscaler.
    
    \item[--] \textbf{Real-time adaptation:} Optimizing the solver performance to enable real-time decision making in large-scale environments.
    
    \item[--] \textbf{Graceful transition strategies:} Developing methods to transition from existing allocations to optimal ones with minimal disruption to running workloads.
\end{itemize}

These future directions would enhance the practical applicability of our convex optimization framework while addressing the complexities of real-world cloud environments. The theoretical foundations established in this work provide a solid basis for these extensions, potentially leading to even greater efficiency gains in cloud resource allocation.

\section{Tasks Assignment}
Shayan contributed to the Approaches, Experiments, and Conclusion sections and Emin implemented the optimization solver and performed the experiments. Amrita contributed to the Introduction and Results sections, and implemented code for the visualizations. Hao-Ting worked on the Statement of Problem and Future Work section.

\bibliography{bibliography.bib}
\bibliographystyle{plain}

\begin{appendices}

\section{Figures}

\setcounter{figure}{0}

\begin{figure}[!h]
    \centering
    \includegraphics[width=0.5\linewidth]{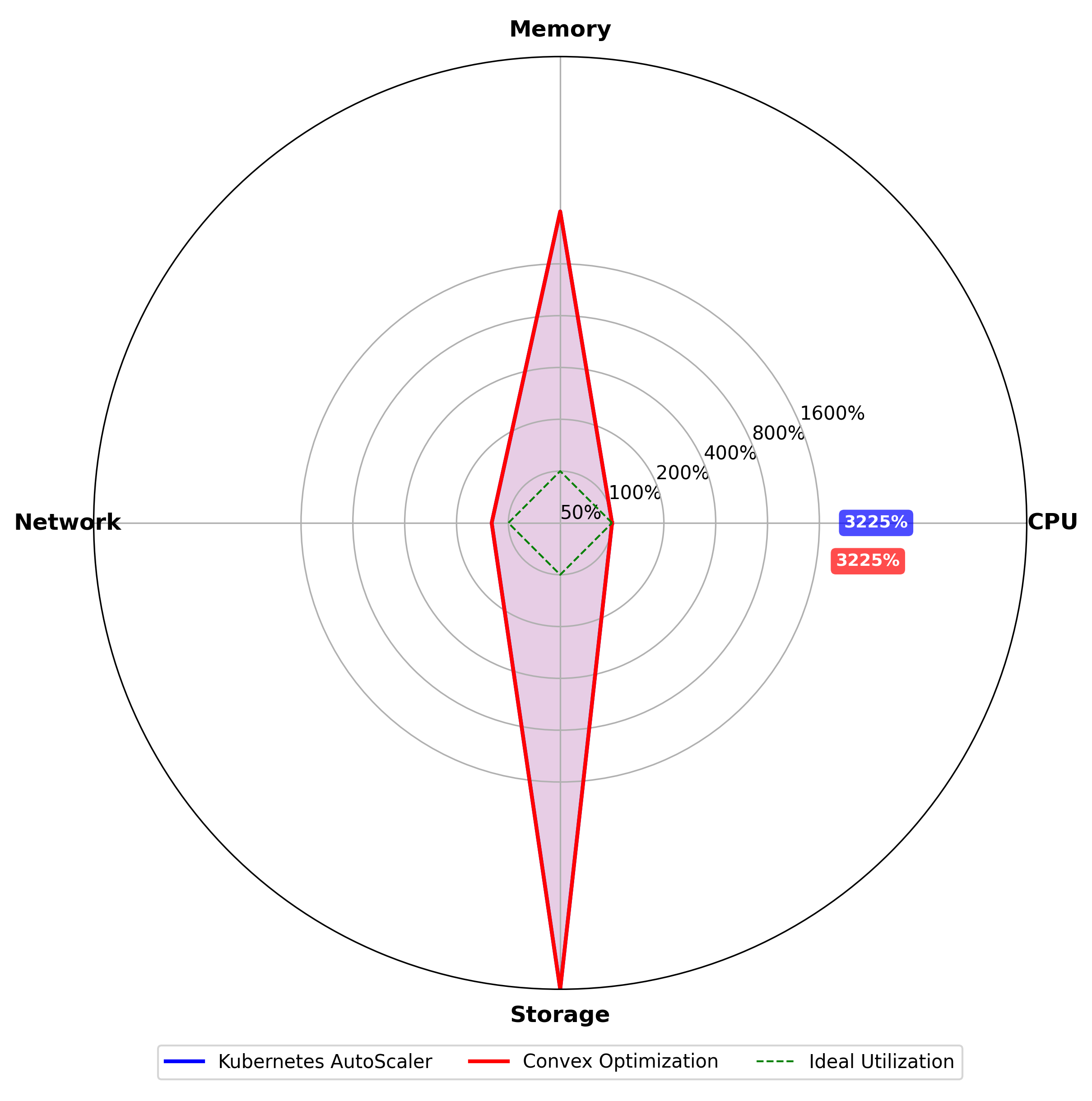}
    \caption{Scenario 1 Resource Radar Graph}
    \label{fig:radar-1}
\end{figure}

\begin{figure}[!h]
    \centering
    \includegraphics[width=0.5\linewidth]{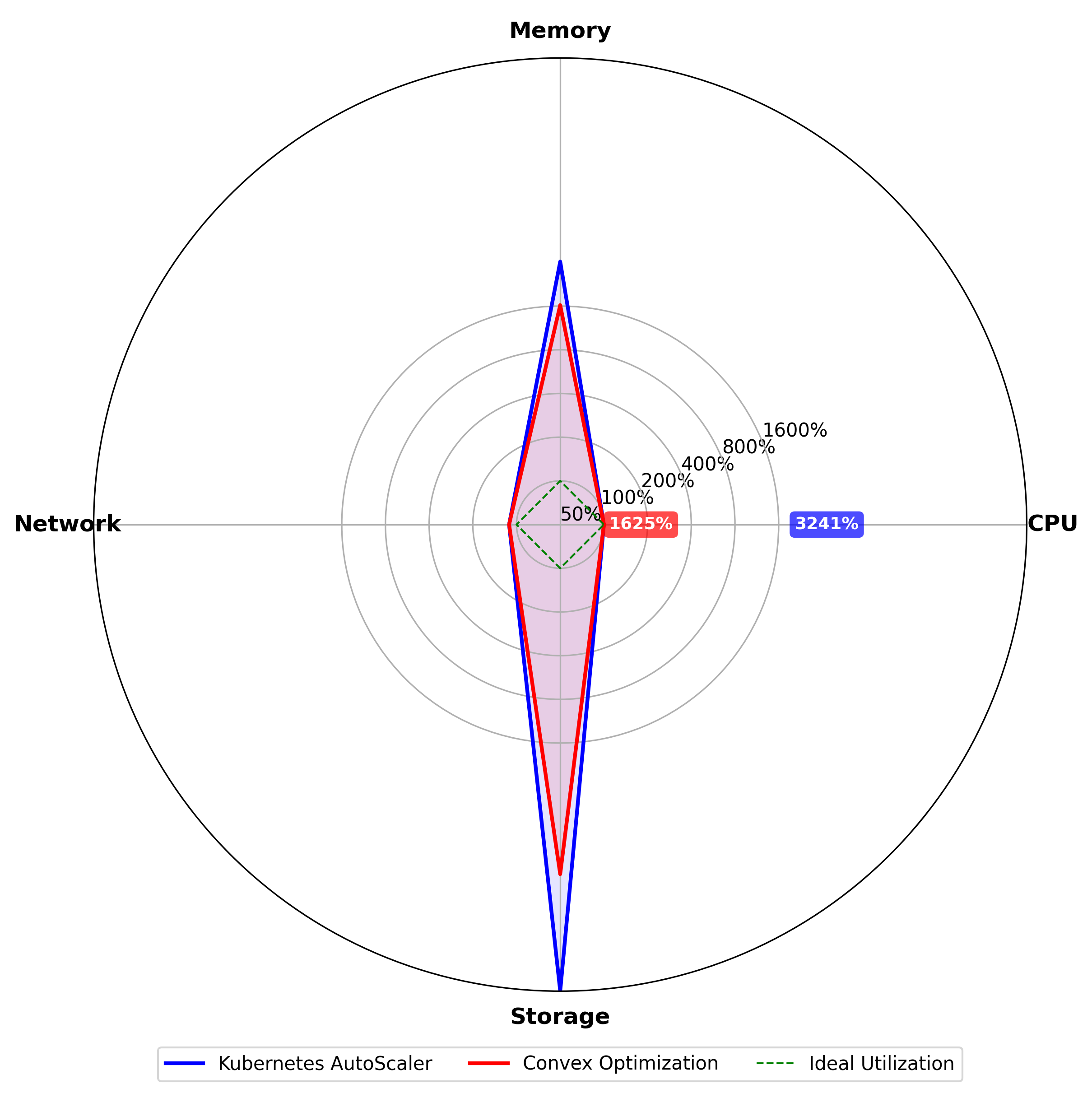}
    \caption{Scenario 2 Resource Radar Graph}
    \label{fig:radar-2}
\end{figure}

\begin{figure}[!h]
    \centering
    \includegraphics[width=0.5\linewidth]{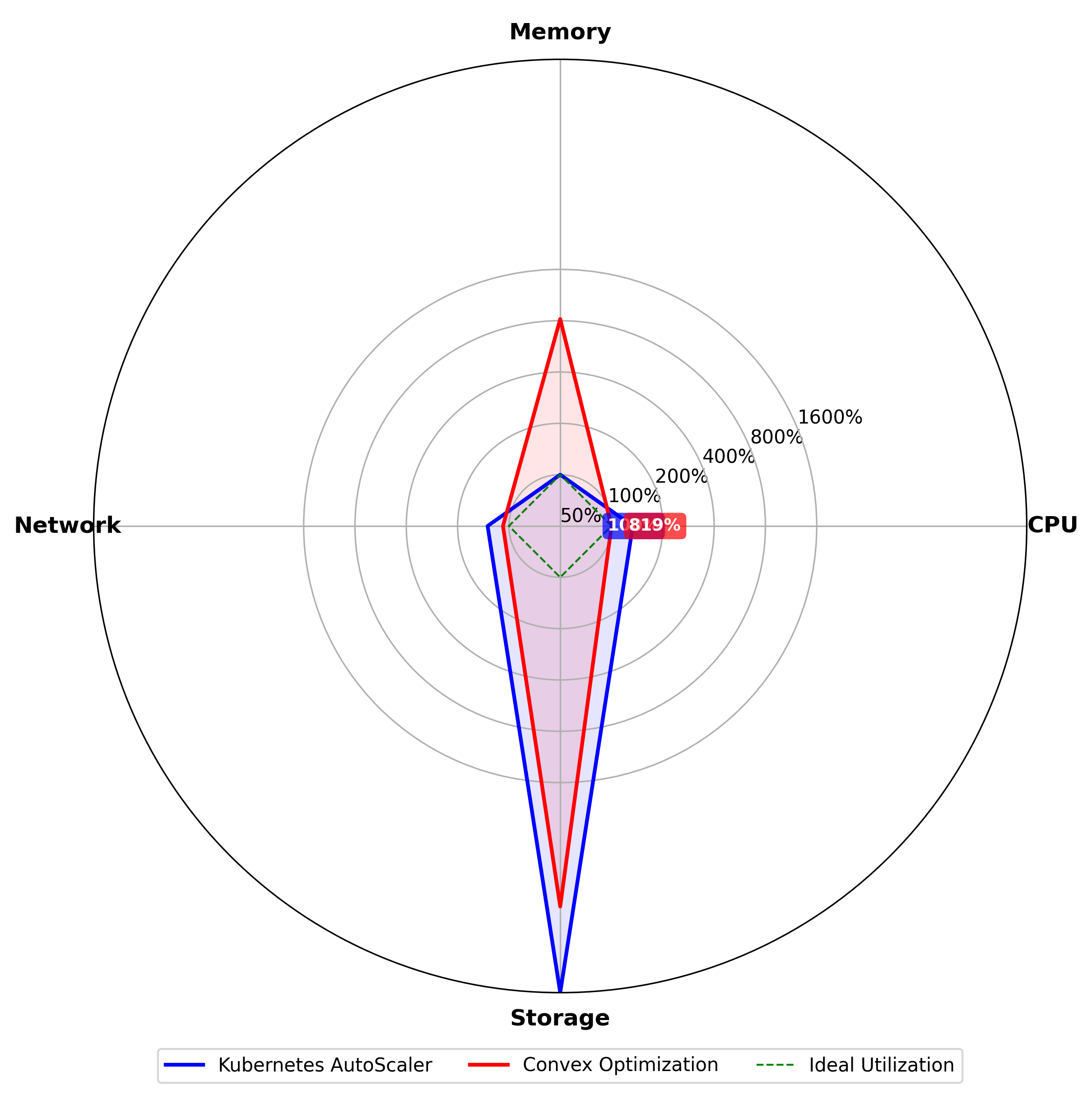}
    \caption{Scenario 3 Resource Radar Graph}
    \label{fig:radar-3}
\end{figure}

\begin{figure}[!h]
    \centering
    \includegraphics[width=0.5\linewidth]{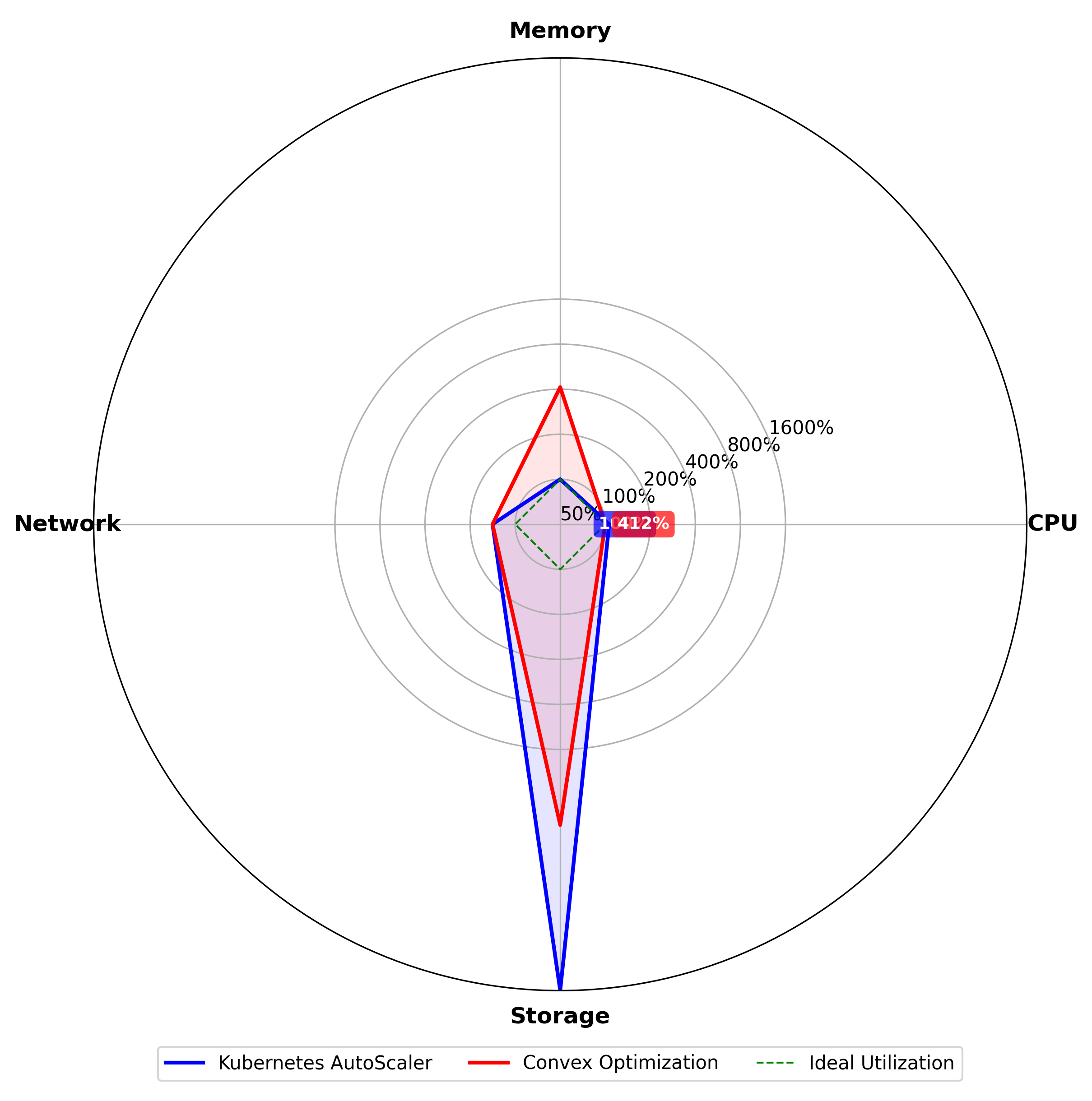}
    \caption{Scenario 4 Resource Radar Graph}
    \label{fig:radar-4}
\end{figure}

\begin{figure}[!h]
    \centering
    \includegraphics[width=0.5\linewidth]{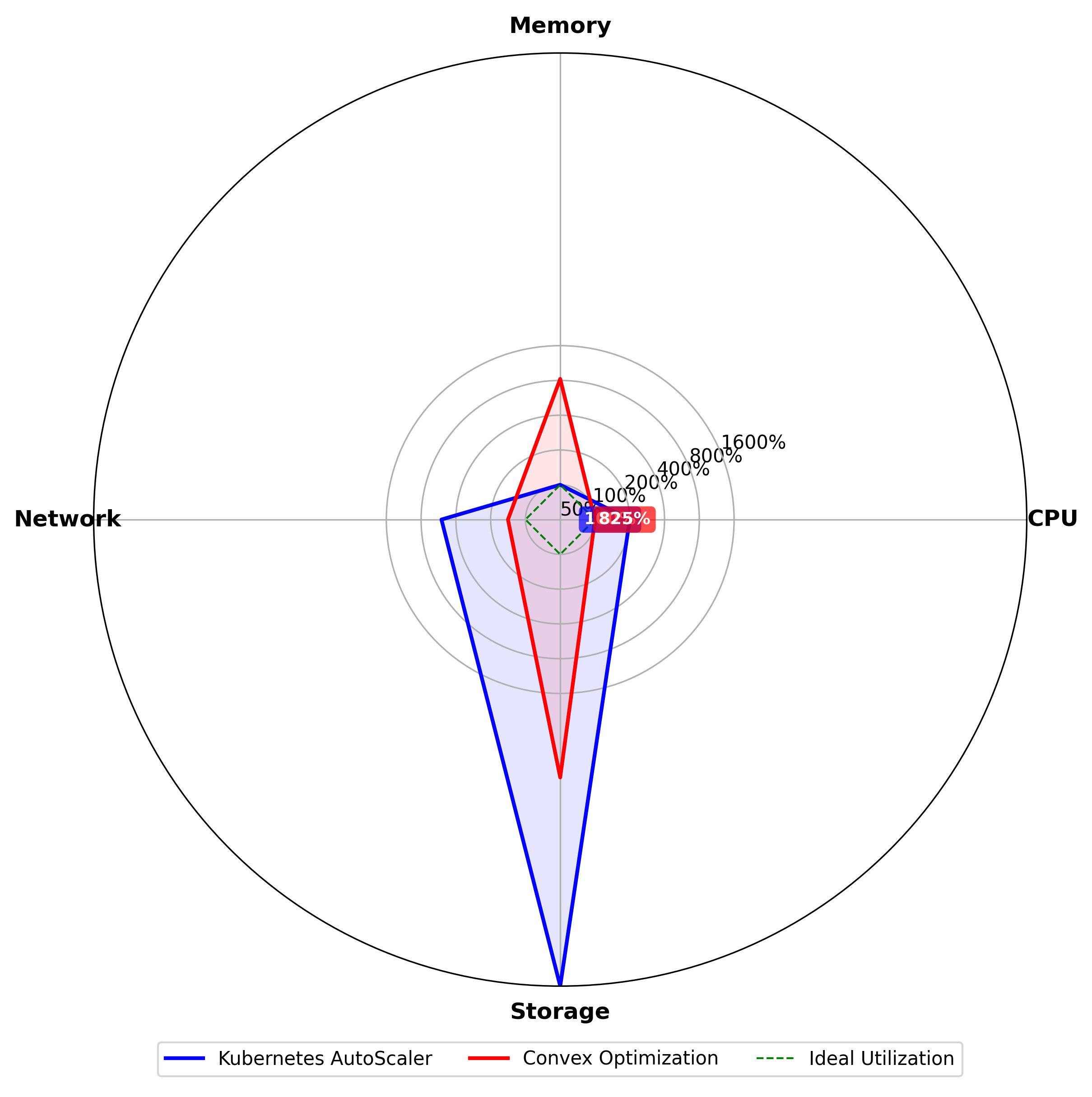}
    \caption{Scenario 5 Resource Radar Graph}
    \label{fig:radar-5}
\end{figure}

\end{appendices}

\end{document}